\documentstyle[11pt,amssymb,epsf,cite]{article}

\makeatletter
\@addtoreset{equation}{section}
\makeatother

\def\ben{\begin{equation}}
\def\een{\end{equation}}

  \let\n=\nu

\let\C=\Chi

\def\nn{\nonumber} \def\bd{\begin{document}} \def\ed{\end{document}}
\def\ds{\documentstyle} \let\fr=\frac \let\bl=\bigl \let\br=\bigr
\let\Br=\Bigr \let\Bl=\Bigl
\let\bm=\bibitem
\let\na=\nabla
\let\pa=\partial \let\ov=\overline
\newcommand{\be}{\begin{equation}}
\newcommand{\ee}{\end{equation}}
\def\ba{\begin{array}}
\def\ea{\end{array}}
\def\ft#1#2{{\textstyle{{\scriptstyle #1}\over {\scriptstyle #2}}}}
\def\fft#1#2{{#1 \over #2}}
\def\del{\partial}
\def\vp{\varphi}
\def\sst#1{{\scriptscriptstyle #1}}
\def\oneone{\rlap 1\mkern4mu{\rm l}}
\def\td{\tilde}
\def\wtd{\widetilde}
\def\ie{\rm i.e.\ }
\def\dalemb#1#2{{\vbox{\hrule height .#2pt
        \hbox{\vrule width.#2pt height#1pt \kern#1pt
                \vrule width.#2pt}
        \hrule height.#2pt}}}
\def\square{\mathord{\dalemb{6.8}{7}\hbox{\hskip1pt}}}
\newcommand{\ho}[1]{$\, ^{#1}$}
\newcommand{\hoch}[1]{$\, ^{#1}$}
\newcommand{\bea}{\begin{eqnarray}}
\newcommand{\eea}{\end{eqnarray}}
\newcommand{\ra}{\rightarrow}
\newcommand{\lra}{\longrightarrow}
\newcommand{\Lra}{\Leftrightarrow}
\newcommand{\ap}{\alpha^\prime}
\newcommand{\bp}{\tilde \beta^\prime}
\newcommand{\tr}{{\rm tr} }
\newcommand{\Tr}{{\rm Tr} }
\def\0{{\sst{(0)}}}
\def\1{{\sst{(1)}}}
\def\2{{\sst{(2)}}}
\def\3{{\sst{(3)}}}
\def\4{{\sst{(4)}}}
\def\5{{\sst{(5)}}}
\def\6{{\sst{(6)}}}
\def\7{{\sst{(7)}}}
\def\8{{\sst{(8)}}}
\def\n{{\sst{(n)}}}
\def\cA{{{\cal A}}}
\def\cF{{{\cal F}}}
\def\tV{\widetilde V}
\def\tW{\widetilde W}
\def\tH{\widetilde H}
\def\tE{\widetilde E}
\def\tF{\widetilde F}
\def\tA{\widetilde A}
\def\im{{{\rm i}}}
\def\tY{{{\wtd Y}}}
\def\ep{{\epsilon}}
\def\vep{{\varepsilon}}
\def\R{\rlap{\rm I}\mkern3mu{\rm R}}
\def\bD{{{\bar D}}}
\def\cD{{\cal D}}
\def\R{\rlap{\rm I}\mkern3mu{\rm R}}
\def\bD{{{\bar D}}}
\def\R{{{\Bbb R}}}
\def\C{{{\Bbb C}}}
\def\H{{{\Bbb H}}}
\def\CP{{{\Bbb C}{\Bbb P}}}
\def\RP{{{\Bbb R}{\Bbb P}}}
\def\Z{{{\Bbb Z}}}
\def\bA{{{\Bbb A}}}
\def\bB{{{\Bbb B}}}
\def\bC{{{\Bbb C}}}
\def\bD{{{\Bbb D}}}
\def\bZ{{{\Bbb Z}}}
\def\Re{{{\frak{Re}}}}
\def\Im{{{\frak{Im}}}}
\def\cosec{{\,\hbox{cosec}\,}}
\def\tX{{{\wtd X}}}

\def\hhtr{Hawking--Hunter--Taylor-Robinson\ }

\newcommand{\mitchell}{\it George P. \& Cynthia W.
Mitchell Institute for Fundamental Physics,\\
Texas A\&M University, College Station, TX 77843-4242, USA}
\newcommand{\tamphys}{\it Center for Theoretical Physics,
Texas A\&M University, College Station, TX 77843, USA}
\newcommand{\umich}{\it Michigan Center for Theoretical Physics,
University of Michigan\\ Ann Arbor, MI 48109, USA}
\newcommand{\upenn}{\it Department of Physics and Astronomy,
University of Pennsylvania\\ Philadelphia,  PA 19104, USA}
\newcommand{\SISSA}{\it  SISSA-ISAS and INFN, Sezione di Trieste\\
Via Beirut 2-4, I-34013, Trieste, Italy}

\newcommand{\ihp}{\it Institut Henri Poincar\'e\\
  11 rue Pierre et Marie Curie, F 75231 Paris Cedex 05}

\newcommand{\damtp}{\it DAMTP and Centre for Theoretical Cosmology, 
Centre for Mathematical Sciences,\\
 University of Cambridge, Wilberforce Road, Cambridge CB3 OWA, UK}
\newcommand{\itp}{\it Institute for Theoretical Physics, University of
California\\ Santa Barbara, CA 93106, USA}

\newcommand{\auth}{{\Large G.W. Gibbons\hoch{\sharp}, M.J. Perry\hoch{\sharp}
and C.N. Pope\hoch{\ddagger} } }

\begin{document}
\begin{flushright}
\hfill{DAMTP-2006-52\ \ \ MIFP-06-16}\\
\hfill{hep-th/0606186}
\end{flushright}


\begin{center}
{ \large {\Large\bf Partition Functions, the Bekenstein Bound and 
Temperature Inversion in Anti-de Sitter Space and its Conformal Boundary 
 }}

\vspace{30pt}
\auth

\vspace{30pt}
{\hoch{\sharp}\damtp}

\vspace{3pt}
{\hoch{\ddagger}\mitchell}

\vspace{30pt}

\underline{ABSTRACT}
\end{center}

   We reformulate the Bekenstein bound as the requirement of positivity of 
the Helmholtz free energy at the minimum
value of the function $L=E- S/(2\pi R)$, where $R$ is some measure
of the size of the system.  The minimum of $L$ occurs at the temperature
$T=1/(2\pi R)$.  In the case of $n$-dimensional anti-de Sitter 
spacetime, the rather poorly defined size $R$ acquires a precise 
definition in terms of 
the AdS radius $l$, with $R=l/(n-2)$.  We previously found that the Bekenstein
bound holds for all known black holes in AdS.  However, in this paper we
show that the Bekenstein bound is not generally valid for free 
quantum fields in AdS, even if one includes the Casimir energy.  
Some other aspects of thermodynamics in 
anti-de Sitter spacetime are briefly touched upon.

{\vfill\leftline{}\vfill \vskip 5pt \footnoterule
{\footnotesize  \hoch{\ddagger} Research supported in part by DOE
grant DE-FG03-95ER40917 and NSF grant INTO3-24081.\vskip  -12pt}}

\pagebreak
\setcounter{page}{1}

\tableofcontents
\addtocontents{toc}{\protect\setcounter{tocdepth}{3}}
\vfill\eject

\section{Introduction} 

    In order to extend to the concept of thermodynamic entropy
gravitating systems, one must include an extra contribution
\be
S_{\rm grav}= { 1 \over 4 G} A\,,
\ee
where $G$ is Newton's constant, and $A$ is the sum of the areas of any 
event horizons within the system.  The generalised second law of 
thermodynamics asserts that the new total entropy cannot decrease.
Some years ago Bekenstein conjectured \cite{bek} that in order for the 
generalised second law to hold, the matter
contribution  to the entropy should satisfy  what has come to be called
the Bekenstein bound,
\ben
S \le  2 \pi R E\,, \label{Bek}
\een
where $S$ is the entropy, $E$ the energy, and $R$ the ``size''
of a  bounded material system capable of being lowered into a black hole.

   The necessity of (\ref{Bek}) for the validity of the generalised
second law was soon called into question \cite{WaldUnruh}.
Nevertheless, Bekenstein's proposed bound has attracted widespread attention
over the succeeding years, not least because of the remarkable feature
that it does not depend upon Newton's constant, and hence it can be
construed as a very general property of all forms of matter, even in
the absence of gravity.

         Another remarkable feature of Bekenstein's suggested bound,
which has not been noticed hitherto, is the existence of a simple,
universal, criterion for the validity of the bound.  It follows
solely from the laws of classical thermodynamics.  Let 
\be
L \equiv  E - \fft{S}{2\pi\, R} = F + \Big(T -\fft1{2\pi\, R}\Big)\, S
\,,\label{ldef}
\ee
where $F=E-TS$ is the free energy,
so that the Bekenstein bound (\ref{Bek}) is equivalent to the statement that
$L \ge 0$.  Since $S=-\del F/\del T$, it follows that $L$ is extremised
if
\bea
0=\fft{\del L}{\del T} &=& \fft{\del F}{\del T} + S + 
  \Big(T -\fft1{2\pi\, R}\Big)\, \fft{\del S}{\del T}\nn\\
&=&   \Big(T -\fft1{2\pi\, R}\Big) \, \fft{\del S}{\del T}\,,
\eea
and hence at the temperature
\be
T= T_L \equiv \fft{1}{2\pi\, R}\,.\label{TLdef}
\ee
The second derivative of $L$ is given by
\be
\fft{\del^2 L}{\del T^2} = \fft{\del S}{\del T} + (T-T_L) 
\fft{\del^2 S}{\del T^2}\,.
\ee
Thus, provided that the specific heat $T\, \del S/\del T$ is positive, 
it follows that there is a unique extremum, and that it is a minimum.
From (\ref{ldef}), the corresponding minimum value of $L$ is
\be
L_{\rm min} = F(T_L)\,,
\ee
\ie the free energy evaluated at the temperature $T_L$ given by
(\ref{TLdef}).  Thus the Bekenstein bound holds for all temperatures
if $F(T_L)$ is non-negative, but it is violated, for some range of
temperatures around $T=T_L$, if $F(T_L)$ is negative.  

   Suppose that the bound is {\it marginally} satisfied, \ie the minimum
value of $L$ is zero.  Then, at that value, the free energy $F$ vanishes.  
At this point, the free energy of the system under consideration equals 
that of the vacuum, and so the system can undergo
a phase transition to the vacuum.  Clearly, this phenomenon is universal;
an example, to be discussed below, being the AdS black
hole and its associated Hawking-Page phase transition.  

   Later in this paper, we shall use the argument above as a criterion
for testing the Bekenstein bound. 

    Since the original statement of the Bekenstein conjecture is
rather vague, a number of questions have to be addressed before its
correctness can be checked. These include

\begin{itemize}

\item What is meant by the entropy $S$, and
what is the ensemble being used?
Is it the microcanonical ensemble, and the
Boltzmann entropy $S_B= \log N(E)$,
where $N(E)$ is the number of states having energy
less than or equal to $E$?\footnote{We prefer to use the
cumulative definition of $N(E)$, which is better behaved,
rather than the often-adopted number of states of energy precisely 
$E$, which jumps
rather erratically with $E$, especially at small $E$ where, as is well known,
the Bekenstein bound is most vulnerable.}  Alternatively, is it
the Gibbs entropy $S_G= -  {\rm Tr} \rho \log \rho$, where
$\rho $ is the normalised density matrix of some ensemble,
such as the grand canonical ensemble?  (Note that these definitions of 
entropy are not necessarily equivalent.  See, for example, \cite{feynman}.)

\item What is meant by the energy $E$?  Does it contain
a contribution from the zero-point energies of the fields,
i.e. the Casimir energy?  Does it contain a contribution
from the box or cavity walls containing the matter?
Should the stress tensor of the  walls of the box
therefore satisfy any restrictions, such as the dominant energy
condition?

\item How is the radius $R$ defined?  What boundary
conditions are to be imposed on the fields at the
boundary of the system?  Are these boundary conditions,
which typically entail divergences,
consistent with some renormalisation scheme ?

\item Should the total number of fields or species be limited
so as to avoid the so-called ``species problem?''
This problem arises if one uses the microcanonical ensemble,
owing to the fact that passing to $N$ identical replicas of the
original system, the Boltzmann entropy
increases as
\ben
S_B \rightarrow S_B + \log N\,,
\een
and so in principle the left-hand side of (\ref{Bek}) could
be made arbitrarily large by taking $N$ sufficiently large.
In fact the species problem does not arise for the Gibbs
entropy of the canonical or grand canonical ensembles. If $Z$
is the partition function for a system with one species of particle, 
then $Z^N$ is the partition function for the same system with $N$ species. 
Since
\bea
E &=& -{\partial \log Z \over  \partial \beta} \longrightarrow N E\,,\nn\\
S_G &=&(1-\beta\, 
        {\partial \over \partial \beta } ) \log Z \longrightarrow N S_G\,,
\label{scale}
\eea
with $\beta = T^{-1}$ being the inverse temperature,
the right-hand side and left-hand side of (\ref{Bek})
scale in an identical fashion\footnote{Note that the often-repeated
statement that the micro-canonical and canonical ensembles
give the same answers at large energies, or equivalently
high temperatures, is not strictly speaking 
correct in the  case of entropy, since the equivalence ignores
sub-leading quantities such as $\log N$}. The correctness of the
scalings (\ref{scale}) is easily checked for a radiation gas
of non-interacting particles.

\item Is it  reasonable at all to use thermodynamic concepts
at the very low temperatures at which the Bekenstein bound is most vulnerable?
Usually one argues that if $T < \tau^{-1}$, where $\tau$ is a typical
relaxation time, which in a finite size cavity cannot be less
than a light crossing time, thermalisation is not possible.
This seems to be an  argument in favour of using  the microcanonical ensemble,
where one is essentially just counting states, \`a la Boltzmann. 
 
\end{itemize}

   Many of the difficulties  raised above may be avoided
if one avoids the technically-demanding
and possibly ill-defined,
situation of a quantum field theory in a
sharply localised  spatially-bounded region in flat Minkowski 
spacetime.\footnote{It is notorious that the sharp confinement of 
quantum fields
leads to many additional divergences, beyond those one encounters
in scattering calculations in infinite space.}  For example, 
one could consider instead a static spacetime
with closed spatial sections, such as the Einstein Static Universe
ESU$_n \equiv {\Bbb R} \times S^{n-1}$,
for which there is no spatial boundary, and hence no need for boundary 
conditions.  Alternatively, one could consider a spacetime with  spatially
non-compact sections, such as those of anti-de Sitter spacetime AdS$_n$,
for which the gravitational
redshift effect is sufficient to confine  a finite-energy system
at non-zero temperature \cite{HawkingPage,AllenFolacciGibbons}.
In this case the boundary conditions are
uncontroversial \cite{BreitenlohnerFreedman,Hawking}, and this makes 
the calculation of zero-point energies straightforward
\cite{AllenDavis,GibbonsNicolai}.
In this respect, as in so many others, it seems that
anti-de Sitter spacetime provides the theorist's perfect adiathermic box,
even to the extent that it permits an infinite uniformly-rotating
platform, and moreover it confines gravitons.
Of course the desire to eliminate the need for boundary
conditions was one of the principal reasons for Einstein's
introduction of the cosmological constant, and his adoption
of $ESU_4$ as a cosmological model. Dowker \cite{Dowker2}
has made a similar point about having
no boundary.  He examined the case  of squashed 3-spheres,
and found that for sufficiently 
large squashing the zero-point contribution
for fermions can be made arbitrarily negative.  Thus, rather remarkably,
there exist temperatures
for which the negative zero-point energy can overwhelm  the 
positive thermal energy, possibly leading to a divergence of the 
ratio $S/E$, but also rendering its sign negative, in gross contradiction to
the Bekenstein conjecture (\ref{Bek}).

   The Einstein Static Universe and anti-de Sitter spacetime
are of course related, in that ESU$_{n-1}=\partial$AdS$_n$,
where $\partial$ denotes the conformal boundary. 
Motivated by the AdS/CFT correspondence, there
has been some recent discussion \cite{KutasovLarsen,Klemm,Dowker,Odd1,Odd2}
of the validity of Bekenstein's bound for
free conformal field theories on ${\Bbb R} \times S^{n-2}$.
Surprisingly, there has been little work or progress on the same problem in
the bulk AdS$_n$ spacetime.  In a recent paper \cite{GibbonsPerryPope}, 
evidence was presented for the general validity
of an AdS Bekenstein bound of the form
\ben
 S < {2 \pi E l \over n-2}\,.\label{AdSBek}
\een
for all known asymptotically anti-de Sitter 
rotating charged black holes, where $l$ is the radius of curvature
of AdS$_n$.  It was pointed out that
the AdS Bekenstein bound is a consequence of the much deeper
and more fundamental conjectured  Cosmic Censorship Bound
for the area $A$ of an apparent  horizon,
\be
E \ge \fft{(n-2) A}{16\pi G\, l}\, \Big[ 
     l\, \Big(\fft{A}{ {\cal A}_{n-2} }\Big)^{-\ft1{n-2}} +
\fft1{l}\,  \Big(\fft{A}{{\cal A}_{n-2} }\Big)^{\ft1{n-2}} \Big]\,.
\label{ccbound}
\ee
Interestingly however, the bound (\ref{AdSBek}) still
does not contain Newton's constant, and so it, like
(\ref{Bek}), may be construed  as a statement
about quantum field theory, or perhaps string theory, 
in a fixed background, namely AdS$_n$. 
Note that by contrast with  (\ref{Bek}),
there is no ambiguity in (\ref{AdSBek})
about the length scale entering the bound.
It is thus a well-defined question to ask whether  (\ref{AdSBek})
is always satisfied for quantum fields, conformal or not, in AdS$_n$.
The main purpose of the paper is to investigate this question,
at the level of free fields.

   In section 2, we briefly describe the relationships between 
one-particle and many-particle partition functions, both for bosons and 
fermions, and we develop some formulae allowing us to calculate 
thermodynamic quantities using zeta functions.  In section 3, we
calculate the energy and entropy for free fields, particularly those
falling into supergravity multiplets, in anti-de Sitter backgrounds in
four, five and seven dimensions.  We find cases where the Bekenstein 
bound is violated, even if the contribution of the Casimir energy is
included.  Section 4 contains a brief discussion, inspired by recent 
work on large-$N$ Yang-Mills theory, of the novel statistics that 
arise when the fields are given by infinite-dimensional matrices
\cite{Sundborg,Polyakov,rams}.  We
find that the Bekenstein bound can be violated in this case too.  The
novel statistics give rise to a maximum Hagedorn-type temperature, which
we calculate.  In section 5, we discuss partition functions for 
conformally-invariant fields on $\R\times S^{n-2}$, or, equivalently, 
AdS$_n$.  All correlation functions exhibit periodicity both in imaginary
time, as a consequence of the non-vanishing temperature, and in real
time because of the periodicity of AdS$_n$.  Such doubly-periodic functions,
provided they have an appropriate analytic structure (e.g. they are
meromorphic), may be expressed in terms of elliptic functions.  If $n$
is odd, the free Green functions have only poles, and in consequence the
partition functions have modular properties under $SL(2,\Z)$ transformations
of the temperature.   If $n$ is even, the Green functions have branch
points even for conformally-invariant fields, and the above arguments fail.  
At the end of the paper, there is an appendix describing how one can 
invert the process of constructing multi-particle partition functions
from single-particle partition functions, by making use of the M\"obius
function and a fermionic generalisation.

\section{Canonical and Grand Canonical Partition Functions}\label{pfsec}

   Suppose that the modes of a free quantum field in a
stationary, axisymmetric
spacetime background ${\cal M}$ are discrete, and have
energies $E$ and angular momentum projections $j$.
One may define a one-particle partition function $Y(\beta, \Omega)$
depending upon the temperature $T=\beta^{-1}$ and chemical potential
$\Omega$ for the angular momentum by
\be
Y(\beta , \Omega)= \sum_{E,\, j} e^{-(\beta E +\alpha j)},\label{Why}
\ee
where
\ben
\alpha =-\beta \Omega.
\een
One may rewrite (\ref{Why}) as
\ben
Y(\beta, \Omega)= {\rm Tr }_{{\cal H}_1}e^{-\beta(\hat H-\Omega \hat j)},
\label{partop}
\een
where $\hat H-\Omega \hat j$ is the quantum mechanical operator
corresponding to the rigidly-rotating Killing field
\ben
{\bf K}={\partial \over \partial t} - \Omega {\partial \over \partial \phi}.
\label{tlkv}
\een
One may regard (\ref{partop}) as an analytic continuation
of a character of the representation of
$e^{-it(\hat H-\Omega \hat j)}$ in the one-particle or ``first quantised''
Hilbert space ${\cal H}_1$ at an imaginary time $t=-i\beta$. 
In computations it is frequently more convenient to
introduce $x^2=e^{-\beta}$ and $y^2=e^{-\alpha}$ and re-write (\ref{Why})
as
\ben
Y(\beta , \Omega)= \sum_{E,j} x^ {2E} y^ {2j}\,,\label{W}
\een
Note that $x$ is always less than one but $y$ may be less than or greater than 
one, depending upon the sign of $\Omega$.
The multi-particle partition function is given by
\ben
Z(\beta, \Omega )= {\rm Tr}_{\cal H} e^{-\beta(\hat H-\Omega \hat j)}\,,
\een
where the trace is taken over the full ``second quantised''
 Hilbert space of
multi-particle states.  For a system of bosons, the multi-particle partition 
function $Z_B(\beta, \Omega)$ is given by
\ben
Z_B(\beta, \Omega)= 
\prod_{E,j} { 1 \over \bigl (1- e^{-\beta (E -\Omega j) }\bigr
 )}\,.
\een
The thermodynamic potential $\Phi(\beta, \Omega)$ is given by
by
\ben
-\beta \Phi(\beta, \Omega)= \log Z_B(\beta, \Omega) =\sum _n { 1\over n}
Y(n \beta,\Omega)\,,
\een
where the summation arises from expanding the logarithm in a Taylor series.

 For a system of fermions the multi-particle partition function
$Z_F(\beta, \Omega) $ is given by
\ben
Z_F(\beta, \Omega)= \prod_{E,j}
\Bigl (1+e^{-\beta (E -\Omega j)}\Bigr )\,.
\een
The thermodynamic potential $\Phi(\beta, \Omega)$ is given by
\ben
-\beta \Phi(\beta, \Omega)= 
     \log Z_F(\beta, \Omega) =-\sum _n { (-1)^{n} \over n}
Y(n\beta,\Omega)\,.
\een
It is clear that all of the information about the thermodynamics
of  a free quantum field theory in a background spacetime is
given by the one-particle partition function.  (See appendix A for a
discussion of how, conversely, the one-particle partition function can 
be recovered from the multi-particle partition function.)

\subsection{The Hamiltonian zeta function}

   Even though  all the relevant spectral information is encoded
into $Y(\beta))$, it is often  more  convenient to
encode it into a zeta function.
 Thus let the Hamiltonian zeta function be defined by 
\ben
\zeta_H(s) = \sum _n  d_n E_n^{-s} = 
{\rm Tr }_{{\cal H}_1}  H^{-s}\,,\label{zetah}
\een
where $H$ is the Hamiltonian acting on the one-particle Hilbert space
${\cal H }_1$, $E_n$ are the energies and $d_n$ the degeneracies of the
states of the system.  Note that the sum (\ref{zetah}) is convergent
provided that $\Re(s)$ is strictly greater than the spatial dimension $N$.
The function $\zeta_H(s)$ can be analytically continued to a meromorphic 
function in the entire
complex plane, with poles only at $s=1,2,\ldots,N$.

   The Hamiltonian zeta function $\zeta_H(s)$
may be obtained from the one-particle partition function
$Y(\beta )$ by a Mellin transform
\ben
\zeta _H(s) = { 1 \over \Gamma(s) } 
\int _0^\infty \beta ^{s-1} Y(\beta ) d \beta \,.
\een
Formally
\ben
\sum_n  d_n  E_n= \zeta _H(-1)\,,
\een 
and so a convenient definition of the Casimir energy  $E_c$ is
\ben
E_c= \ft12 (-1)^F \zeta_H(-1)\,,\label{Cas}
\een
where $F$ is $0$ for bosons and $1$ for fermions.

   Using the identity
\ben
e^{-\beta} = { 1 \over 2 \pi i} \int ^{c+i\infty}_{c-i\infty} 
\beta^{-s}\,  \Gamma(s) ds \,,
\een 
where $c>0$, 
one finds that the ``blind'' grand canonical partition function
(\ie with $\Omega=0$)  for bosons may be expressed as 
\ben
\log Z_B(\beta)= 
{ 1 \over 2 \pi i} \int ^{\gamma +i\infty}_{\gamma -i\infty} 
\beta^{-s}\,  \Gamma(s) \zeta(s+1) \zeta_H(s) ds \,,
\een
where $\zeta(s)$ is the Riemann zeta function, 
while for fermions 
\ben
\log Z_F(\beta)= 
{ 1 \over 2 \pi i} \int ^{ \gamma +i\infty}_{\gamma -i\infty} 
\beta^{-s} \, 
\Gamma(s) ( 1- 2^{-s}  ) \zeta(s+1) \zeta_H(s) ds \,. 
\een
In these formulae one must take $\gamma$ to be greater than the
spatial dimension $N$, so that the order of integration and summations
may be freely interchanged.

    From $E= -\del \log Z/\del\beta$, it follows that 
for bosons
\ben
E_B= \sum_n {d_n E_n \over e^{\beta E_n}-1} =
 { 1 \over 2 \pi i} \int ^{\gamma +i\infty}_{\gamma-i\infty} 
\beta^{-s}\,  \Gamma(s) \zeta(s) \zeta _H (s-1)  ds\,,\label{bint}
\een
while for fermions
\ben
E_F = \sum_n {d_n E_n \over e^{\beta E_n}+1} =
 { 1 \over 2 \pi i} \int ^{\gamma +i\infty}_{\gamma -i\infty} 
\beta^{-s}\,  \Gamma(s) (1-  2^{1-s})
 \zeta(s) \zeta _H (s-1)  ds \,.\label{fint}
\een
In deriving these one uses $s\Gamma(s)=\Gamma(s+1)$, and then changes 
variable according to $s\rightarrow s-1$.  This implies that $\gamma$ must
now be taken to be greater than $N+1$.

\section{Entropy and Energy in Bulk AdS Spacetimes}

   Fields in anti-de Sitter spacetime are taken to satisfy reflecting boundary
conditions at infinity.  An $n$-dimensional AdS spacetime, satisfying 
$R_{\mu\nu}=- (n-1) l^{-2} g_{\mu\nu}$, thus provides a perfect realisation of
a closed spherical adiathermic box of radius $l$, which can be used in 
order to study 
the thermodynamics of isolated closed systems.  In particular, it allows
one to give meaning to the otherwise somewhat ill-defined notion of
a thermodynamic system of radius $l$ that is called for in the 
formulation of the Bekenstein bound, which asserts that the energy of
such a system is bounded by
\be
E\ge \fft{(n-2) S}{2\pi l}\,,\label{bekbound}
\ee
where $S$ is the entropy. 

   In this section, we calculate the partition functions for free fields in
anti-de Sitter spacetimes, and use these to study the Bekenstein bound
in the idealised AdS$_n$ ``laboratory.''  Before doing this, we begin with
a discussion of the high-temperature limit.

\subsection{High-temperature expansions and Tolman redshifting}

   At high temperature, free quantum fields in AdS$_n$ behave like a
radiation gas whose local temperature redshifts according to Tolman's
well-known formula \cite{tolman}
\be
T_{\rm local} \, \sqrt{-g_{00}} = \fft1{\beta}\,,
\ee
where $T=\beta^{-1}$ is the global temperature as measured by an observer
at rest at the origin.
When allowance is made for the redshifting, the total energy is given by
\be
E = \sigma\, \beta^{-n}\, \int (-g_{00})^{(n-1)/2} \, \sqrt{\det(g_{ij})}\,
   d^{n-1} x = \sigma\, \beta^{-n}\, V_{\rm eff}\,,
\ee
where 
\be
V_{\rm eff} \equiv  \int (-g_{00})^{(n-1)/2} \, \sqrt{\det(g_{ij})}\,
   d^{n-1} x\,,
\ee
and $g_{ij}$ denotes the spatial $(n-1)$-metric (\ie the AdS$_n$ metric 
$g_{\mu\nu}$ with its indices restricted to the spatial directions).
As pointed out by Hawking and Page \cite{HawkingPage}, the effective
volume $V_{\rm eff}$ is finite, and in fact given by $V_{\rm eff}= 
\ft12 {\cal A}_{n-1}$, where ${\cal A}_m$ is the volume of the unit
$m$-sphere.  The quantity $\sigma$ is the generalisation of the
Stefan-Boltzmann constant, and is given by
\bea
{\rm Bosons}:\qquad && \sigma= (2\pi)^{1-n}\, 
{\cal A}_{n-2} \, \zeta(n)\, (n-1)!\,,\nn\\
{\rm Fermions}: \qquad && \sigma= (2\pi)^{1-n}\, 
{\cal A}_{n-2}\, (1-2^{-n+1})\, \zeta(n)\,
   (n-1)!\,. 
\eea

    At high temperature, the entropy $S$ and free energy $F$ are related
to the total energy $E$ by
\be
S \longrightarrow \fft{n}{n-1}\, \fft{E}{T}\,,\qquad
F \longrightarrow \fft1{n-1}\, E\,.
\ee
Thus
\be
\log Z \longrightarrow \fft{\sigma\, V_{\rm eff}}{(n-1)\, \beta^{n-1}}\,.
\ee
This same result may be derived microscopically either by considering the
point-split non-zero-temperature Green function \cite{AllenFolacciGibbons}, 
or, more economically, by noting that at high temperature $Y(\beta) \sim
Y_0/\beta^{n-1}$, and $\log Z\sim \zeta(n-1)\, Y_0/\beta^{n-1}$ for 
bosons, or $\log Z\sim (1-2^{1-n}) \zeta(n-1)\, Y_0/\beta^{n-1}$ for
fermions.  The coefficient $Y_0$ determines the density of states at 
high energy in a cavity of effective volume $V_{\rm eff}$, and is well
known to be universal, independent of the shape of the cavity.  One readily
checks that this agrees with the radiation-gas approximation.

\subsection{Entropy bound in AdS$_4$}

\subsubsection{AdS$_4$ Partition functions in the canonical ensemble}

   Massless fields in AdS$_4$ are characterised by certain unitary irreducible
representations $D(E_0,s)$ of $SO(2,3)$, where $E_0$ is the lowest energy
and $s$ is the spin.  In general, for spin $s\ge \ft12$, the massless 
representations
correspond to taking $E_0=s+1$.  For massless conformally-invariant 
scalars there are two representations, namely $D(1,0)$ and $D(2,0)$.  If we
normalise the scale by taking $l=1$, then the energies $E_{n,j}$ and 
degeneracies $d_{n,j}$ are given by
\be
E_{n,j}= n+j+1\,,\qquad d_{n,j}= 2j+1\,,\qquad \hbox{where}\qquad
    n\ge 0\,,\qquad j\ge s\,,
\ee
with $n$ and $j$ increasing in integer steps.
 
    From these expressions, and taking the angular velocity to be zero
for now, one finds that the single-particle partition
functions $Y_{(E_0,s)}(\beta)$ for free fields in the $D(E_0,s)$ 
representation are as follows:
\bea
Y_{(1,0)} &=& \fft{e^{2\beta}}{(e^\beta -1)^3} \,,\qquad
Y_{(2,0)} = \fft{e^{\beta}}{(e^\beta -1)^3} \,,\nn\\
Y_{(s+1,s)}&=& \fft{e^{(1-s)\beta}\, [(2s+1) e^\beta + 1-2s]}{(e^\beta-1)^3}\,,
\qquad s=\ft12,1,\ft32,2,\ldots\label{yads4}
\eea
We can also consider the singleton ``Di'' and ``Rac'' representations
$D(1,\ft12)$ and $D(\ft12,0)$ respectively, for which one finds the
single-particle partition functions
\be
Y_{(1,\ft12)}(\beta) = \fft{2}{(e^\beta -1)^2}\,,\qquad
 Y_{(\ft12,0)}(\beta) = 
     \fft{e^{\ft12\beta}\, (e^\beta +1)}{(e^\beta-1)^2}\,.\label{ysingads4}
\ee

The multi-particle partition functions are then calculated using the
expressions (\ref{bosonmp}) and (\ref{fermionmp}) given in appendix A.
One then has the expressions
\be
E(\beta)= -\fft{\del}{\del\beta} \log Z\,,\qquad S(\beta)= \log Z - 
            \beta\, \fft{\del}{\del\beta} \log Z\label{ESdef}
\ee
for the free energy and the entropy of the system.
    
   The Bekenstein bound (\ref{bekbound}), applied to the case of AdS$_4$ 
with $l=1$, asserts that 
\be
L(\beta)\equiv E(\beta) - \fft{S(\beta)}{\pi} \ge 0\,.
\ee
As we showed in the introduction, and can be seen also 
from (\ref{ESdef}), $L(\beta)$ attains its minimum
value when $\beta=\pi$, implying
\be
L_{\rm min} = L(\pi) = -\fft{1}{\pi}\, \log Z\,,\label{fmin}
\ee
and so it is here, at temperature $T=1/\beta=1/\pi$ that one obtains the most
stringent test of the validity of the Bekenstein bound.

   Whilst the Bekenstein bound is clearly satisfied in the high-temperature
regime, where $E\sim T^4$ and $S\sim T^3$, it is not so easy to check the
bound analytically at $T=1/\pi$.  However, it is straightforward to perform
the summations in (\ref{bosonmp}) and (\ref{fermionmp}) numerically to
the required degree of accuracy.  Some explicit results for fields in the
$D(E_0,s)$ representations are as follows:
\bea
L_{\rm min}(1,0) &=& -0.0160124\,,\qquad
L_{\rm min}(2,0) = -0.00067922\,,\nn\\
L_{\rm min}(2,1) &=& -0.0020083\,,\qquad
L_{\rm min}(3,2) = -0.000142841\,,\nn\\
L_{\rm min}(\ft32,\ft12) &=& -0.00650369\,,\qquad
L_{\rm min}(\ft52,\ft32) = -0.000552029\,.\label{fmin1}
\eea
Additionally, for the Di and Rac singletons we find
\be
L_{\rm min}(1,\ft12) = -0.029472\,,\qquad 
L_{\rm min}(\ft12,0) = - 0.0834542\,.\label{diracfmin}
\ee
As can be seen, they are all negative, which would be in contradiction to
the Bekenstein bound.  A representative plot of $L(\beta)$ for a scalar
in the $D(2,0)$ representation is given in Figure 1 below.

\bigskip

\begin{figure}[ht]
\epsfxsize=3.4truein
\leavevmode\centering
\epsfbox{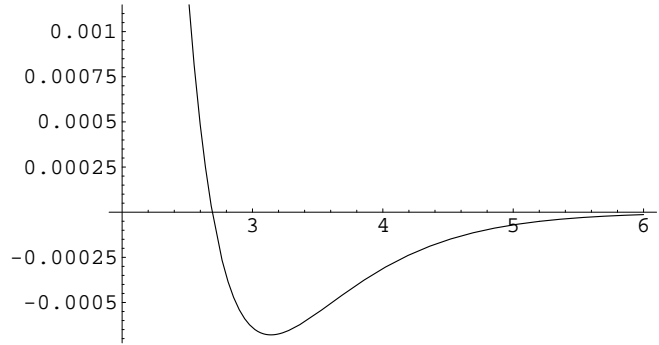}
\caption{{\it A plot of $L=E-S/\pi$ for a scalar field in the $D(2,0)$ 
representation
of $SO(3,2)$ in AdS$_4$, as a function of $\beta=1/T$.  The function 
attains its minimum at $\beta=1/T=\pi$.  It is positive at high 
temperature, but it is negative at sufficiently low temperature.}}
\end{figure}

  The situation changes somewhat if we include the Casimir energies in the 
calculations.
Evaluating these using zeta-function regularisation, as in (\ref{Cas}), 
one finds the additional contributions
\bea
E_c(1,0) &=& \ft1{480}\,,\qquad E_c(2,0) = \ft1{480}\,,\nn\\
E_c(2,1) &=& \ft{11}{240}\,,\qquad E_c(3,2) = \ft{401}{240}\,,\nn\\
E_c(\ft32,\ft12) &=& \ft{17}{1920}\,,\qquad 
   E_c(\ft52,\ft32) = -\ft{863}{1920}\,.\label{ads4casimirs}
\eea
For the singletons, we have
\be
E_c(1,\ft12) =0\,,\qquad E_c(\ft12,0) =0\,.
\ee
With these included, the results for $f_{\rm min}$ in (\ref{fmin1})
are replaced by $\tilde f_{\rm min}$, given by
\bea
\td L_{\rm min}(1,0) &=& -0.0139291\,,\qquad
\td L_{\rm min}(2,0) = +0.00140411\,,\nn\\
\td L_{\rm min}(2,1) &=& +0.043825\,,\qquad
\td L_{\rm min}(3,2) = +1.67069\,,\nn\\
\td L_{\rm min}(\ft32,\ft12) &=& +0.00235048\,,\qquad
L_{\rm min}(\ft52,\ft32) = -0.450031\,,\label{fmin11}
\eea
with the minima for the Di and Rac singletons unchanged from 
(\ref{diracfmin}).  Although in some cases the Casimir energy reverses 
the sign, in others the violation of the Bekenstein bound persists. 

\subsubsection{Entropy bounds in the microcanonical ensemble}

   One can also study the Bekenstein bound in the microcanonical
ensemble, where the energy is held fixed.  To do this, one takes an 
inverse Laplace transform of the expression
\be
Y(\beta) = \int_0^\infty \rho(E)\, e^{-\beta E} dE
\ee
for the single-particle partition function, in order to obtain the 
expression for the density of states $\rho(E)$.  From the expressions in
(\ref{yads4}), we therefore obtain
\bea
\rho_{(1,0)}(E) &=& \ft12 \sum_{n\ge 1} \delta(E-n)\, n(n+1)\,,\nn\\
\rho_{(2,0)}(E) &=& \ft12 \sum_{n\ge 2} \delta(E-n)\, n(n-1)\,,\nn\\
\rho_{(s+1,s)}(E) &=& \sum_{n\ge s} \delta(E-n)\, (n^2-s^2)\,,\qquad
           s=\ft12,1,\ft32, 2,\ldots \,.\label{rhoads4}
\eea
For the Di and Rac singletons we find from (\ref{ysingads4})
\be
\rho_{(1,\ft12)}(E) = \sum_{n\ge 1} \delta(E-n)\, n\,,\qquad
\rho_{(\ft12,0)}(E) = \sum_{n\ge \ft12} \delta(E-n)\, n\,.\label{rhoads42}
\ee

  From these expressions, one can integrate to obtain the total number
of states $N(E)$ with energies less than or equal to $E$.  Thus one has
\bea
N_{(1,0)}(E)&=& \ft12 \sum_{n=1}^E n(n+1) = \ft16 E(E+1)(E+2)\,,\nn\\
N_{(2,0)}(E)&=& \ft12 \sum_{n=2}^E n(n-1) = \ft16 E(E^2-1)\,,
\label{canfields}\\
N_{(s+1,s)}(E)&=& \sum_{n=s}^E (n^2-s^2) = \ft16 (E-s)(E+1-s)(2E+4s+1)\,,
           \quad s=\ft12,1,\ft32, 2,\ldots \,,\nn
\eea
and for the singletons
\be
N_{(1,\ft12)}(E) = \sum_{n=1}^E n = \ft12 E(E+1)\,,\qquad
N_{(\ft12,0)}(E) = \sum_{n= \ft12}^E n = \ft12 (E+\ft12)^2\,.
\label{Nsingletons}
\ee

   At large $E$, the entropy in the microcanonical ensemble, which is 
defined as $S=\log N(E)$, is of the form
\be
S \sim 3 \log E
\ee
for all the fields in (\ref{canfields}).
Thus the ratio $S/(\pi E)$ at large $E$ approaches $3/(\pi E)\, \log E$, 
which tends to zero, implying that the Bekenstein bound is satisfied in
this regime.  Similarly, for the singletons, we have $S\sim 2 \log E$, and
again the Bekenstein bound is satisfied at large $E$.

\subsection{Partition functions and entropy bounds in AdS$_5$}

   With the angular velocities set to zero, the resulting ``blind'' 
single-particle partition
functions for the cases of $(j_1,j_2)$ and $(j_1,0)$ representations of the 
little group $SO(4)$ may be found in \cite{dolan}. These are:
\bea
(j_1,j_2):&& Y = \fft{s^{j_1 + j_2 +2} }{(1-s)^4}\, [(2j_1+1)(2j_2+1) - 
    4 s j_1 j_2]\,,\nn\\
(j_1,0): && Y= \fft{s^{j_1+1} }{(1-s)^3}\, [2j_1+1 - s (2j_1-1)]\,,
\eea
where $s\equiv e^{-\beta}$.  
The first line is for massless representations, and the second is for
doubletons.  By taking the inverse Laplace transform, we find the
corresponding densities of states:
\bea
(j_1,j_2): && \rho(E) = \ft16\sum_{m\ge j_1 + j_2 + 2} \delta(E-m)\, 
    (m-j_1-j_2-1)(m-j_1-j_2)\times\nn\\
&& \qquad\qquad \qquad [m(1+2j_1+2j_2) + 1+ j_1 + j_2 - 2 j_1^2 - 2
    j_2^2 + 8 j_1 j_2]\,,\nn\\
(j_1,0):&& \rho(E)= \sum_{m\ge j_1+1} \delta(E-m) (m^2-j_1^2)\,.
\eea

   The above calculation allows us to read off the energies $E$ and 
degeneracies $d$ for a ``standard'' massless field\footnote{That is, a 
massless field with $E_0=2+j_1+j_2$.}  in the 
$(j_1,j_2)$ representation of the $SO(4)$ little group:
\bea
E_k &=& 2 + j_1 + j_2 + k\,,\qquad \qquad k=0,1,2,\ldots\nn\\
d_k &=& \ft16(k+1)(k+2)[(1+2j_1+2j_2)k + 3(2j_1+1)(2j_2+1)]\,.\label{edegen}
\eea
For massless fields whose $E_0$ value exceeds the minimum $2+j_1+j_2$, 
one just takes $E_k=E_0 + k$, with $d_k$ again given by (\ref{edegen}).

    The fields in the $N=8$ massless supergravity multiplet are
characterised by their $SO(4)$ little-group representations $(j_1,j_2)$,
their lowest energies $E_0$, and their $SU(4)$ gauge-group representations.  
These are given in Table 1 below:

\bigskip\bigskip
\centerline{
\begin{tabular}{|c|c|c|c|} \hline
Field & $SO(4)$ Rep. & $E_0$  &  $SU(4)$ Rep.\\ \hline \hline
Scalar & $(0,0)$ & 2 & $20'$ \\ 
 & $(0,0)$ & 3 &  $10$ + $\overline{10}$  \\
 & $(0,0)$ & 4 & 1 + 1 \\  \hline
Vector & $(\ft12,\ft12)$ & 3 & $15$\\ \hline
A/sym. tensor & $(1,0)$ + $(0,1)$ & 3 & 6 + 6 \\ \hline
Spin 2 & $(1,1)$ & 4 & 1 \\ \hline
Spin $\ft12$ & $(\ft12,0)$ + $(0,\ft12)$ & $\ft52$ & $20$ + 
                    $\overline{20}$ \\ 
 & $(\ft12,0)$ + $(0,\ft12)$ & $\ft72$ & $4$ + $\overline{4}$ \\ \hline
Spin $\ft32$ & $(1,\ft12)$ + $(\ft12,1)$ & $\ft72$ 
      & $4$ + $\overline{4}$ \\ \hline
\end{tabular}}
\bigskip

\noindent{\bf Table 1:}  Lowest energies $E_0$ for the 
$N=8$ massless supermultiplet
\bigskip\bigskip

   We see that all fields except the $10$, $\overline{10}$, 1, 1 of scalars,
and the $4$ and $\overline{4}$ of spin $\ft12$ fields, have the minimum value
$E_0=2+j_1+j_2$ for their lowest-energy.  From the expressions in 
(\ref{edegen}), we can read off the energies and degeneracies for 
the fields in the massless supermultiplet, and then use these to calculate
the Casimir energy for each field.  The results are given in Table 2 
below:

\bigskip\bigskip
\centerline{
\begin{tabular}{|c|c|c|c|c|} \hline
Field & $SU(4)$ Rep. & $E$ & Degeneracy & Casimir Energy\\ \hline \hline
Scalar & $20'$ & $k+2$ & $\ft16 (k+1)(k+2)(k+3)$ & 0\\ 
 & $10$ + $\overline{10}$ & $k+3$ & $\ft16 (k+1)(k+2)(k+3)$ & $-\ft1{480}$ \\
 & 1 + 1 & $k+4$ & $\ft16 (k+1)(k+2)(k+3)$  & $\ft3{80}$ \\  \hline
Vector & $15$ & $k+3$ & $\ft12(k+1)(k+2)(k+4)$ & $-\ft{11}{240}$ \\ \hline
A/sym. tensor & 6 + 6 & $k+3$ & $\ft12 (k+1)(k+2)(k+3)$ & 
                                        $-\ft1{160}$\\ \hline
Spin 2 & 1 & $k+4$ & $\ft16 (k+1)(k+2)(5k+27)$ & $-\fft{553}{240}$ \\ \hline
Spin $\ft12$ & $20$ + $\overline{20}$ & $k+\ft52$ &
           $\ft13(k+1)(k+2)(k+3)$ & $-\ft{17}{3840}$ \\ 
 & $4$ + $\overline{4}$ & $k+\ft72$ & $\ft13(k+1)(k+2)(k+3)$ &
          $\ft{29}{3840}$ \\ \hline
Spin $\ft32$ & $4$ + $\overline{4}$ & $k+\ft72$ & 
           $\ft13(k+1)(k+2)(2k+9)$ & $\ft{141}{320}$ \\ \hline
\end{tabular}}
\bigskip

\noindent{\bf Table 2:}  Energies and degeneracies for each field in the $N=8$ 
massless supermultiplet.   In each case $k=0,1,2,3\ldots$.  The Casimir
energy, calculated using zeta-function regularisation, is given 
{\it per field}.
\bigskip\bigskip

   If we now add up the contributions to the Casimir energy from each field,
we obtain the total
\be
E_c = \ft{3}{8}
\ee
for the entire massless supermultiplet.  This can be compared with the 
Casimir energy $\ft3{16}$ for the $N=4$ super-Yang-Mills multiplet in
the boundary conformal field theory.

   The total single-particle partition functions for the bosonic and
fermionic fields of the $N=8$ supermultiplet are given by
\bea
Y_{\rm boson}(\beta) &=& \fft{4(e^{5\beta}-1)(e^\beta + e^{-\beta} + 6)}{
        (e^\beta -1)^4}\,,\nn\\
Y_{\rm fermion}(\beta) &=& \fft{16(e^{5\beta}-1)
     (e^{\ft12}\beta + e^{-\ft12\beta})}{(e^\beta -1)^4}\,.
\eea
The total single-particle partition function for the entire 
$N=8$ supermultiplet is then given by
\be
Y_{\rm tot}(\beta) = \fft{4(5- e^{-\beta})}{(e^{\ft12\beta} -1)^4}\,.
\ee

   Calculating the total free energies and entropies for the bosonic
and fermionic sectors from their respective multi-particle partition
functions, and then summing these to get the total free energy $E_{\rm
tot}(\beta)$ and entropy $S_{\rm tot}(\beta)$ for the $N=8$
supermultiplet, we can examine the Bekenstein bound.  This asserts
that
\be
L(\beta) \equiv E_{\rm tot}(\beta) - \fft{3 S_{\rm tot}(\beta)}{2\pi}
  \ge 0\,.
\ee
It is easily seen that the lowest value for $L(\beta)$ will occur at
$\beta= 2\pi/3$, and for this value we find
\be
L_{\rm min} = -0.838186\,.
\ee
Including the Casimir contribution $E_c= 3/8$ is insufficient to 
outweigh this, and so there is a range of temperatures corresponding
to approximately to
\be
1.84884 < \beta < 2.744356
\ee
within which the Bekenstein bound is violated.

\subsection{Partition functions and entropy bounds in AdS$_7$}

   First, we need to determine the partition functions for massless
fields in AdS$_7$.  We use equation (3.25) in \cite{dolan} for this, with
the simplifying specialisation to the ``blind'' case where the
$x_i$ factors parameterising 
the chemical potentials for angular momenta are all set to 1.
 
   In the dominant highest-weight labelling $(\ell_1,\ell_2,\ell_3)$
of $SO(6)$ little-group representations, the partition function,
as defined in \cite{dolan}, is given by
\be
D^{(6)}_{[\ell_1 +4; \ell_1, \ell_2,\ell_3]}(s,1) =
  s^{\ell_1+4}\, [\chi^{(6)}_{(\ell_1,\ell_2,\ell_3)} - s\, 
          \chi^{(6)}_{(\ell_1 -1, \ell_2,\ell_3)} ]\,
           P^{(6)}(s,1)\,,\label{ads7part}
\ee
where
\bea
\chi^{(6)}_{(\ell_1,\ell_2,\ell_3)}
&=& \ft1{12}(1+\ell_1-\ell_2)(1+\ell_2-\ell_3)(1+\ell_2+\ell_3)
 (2+\ell_1-\ell_3)\times\nn\\
&& \qquad \qquad\times (2+\ell_1+\ell_3)(3+\ell_1+\ell_2)\,,\nn\\
P^{(6)}(s,1) &=& \fft{1}{(1-s)^6}\,,
\eea
and $s\equiv e^{-\beta}$.  

   The relation between dominant highest-weight labels and Dynkin
labels for representations of $SO(2n)$ is that with $\underline{\ell} = 
\sum \ell_i \underline{e_i}$ in the orthonormal basis $\underline{e_i}$
for $\R^n$, one has the simple roots 
\bea
\underline{\alpha}_i &=& \underline{e}_i - \underline{e}_{i+1}\,,\qquad
   1\le i \le n-1\,,\nn\\
\underline{\alpha}_n &=& \underline{e}_{n-1} + \underline{e}_n\,.
\eea
One then gets the Dynkin labels as the dot products of $\underline{\ell}$
with the simple root vectors $\underline{\alpha}_i$.  For the case of
interest to us here, \ie $SO(6)$, we shall use the Dynkin 
labelling for $SU(4)$, which means one orders the labels as
\be
(\underline{\alpha}_2\cdot \underline{\ell}, 
\underline{\alpha}_1\cdot \underline{\ell}, 
\underline{\alpha}_3\cdot \underline{\ell})\,.
\ee
In other words, the
$SU(4)$ Dynkin label for a highest-weight representation 
$(\ell_1,\ell_2,\ell_3)$ is written as
\be
(\ell_2-\ell_3, \ell_1-\ell_2, \ell_2 + \ell_3)\,.
\ee

   From \cite{guntak}, the AdS$_7$ massless supergravity 
multiplet is as given in Table 3 below.  We list the fields, their 
$SU(4)$ little-group Dynkin labels, their corresponding highest-weight 
labels, their $E_0$ values and their $SO(5)$ R-symmetry representations

\bigskip\bigskip
\centerline{
\begin{tabular}{|c|c|c|c|c|} \hline
Field & $SU(4)$ Dynkin & $SU(4)$ HW & $E_0$ & $SO(5)$ rep.\\ \hline \hline
Scalar & $(0,0,0)$ & $(0,0,0)$ & 4 & 14 \\ \hline
Spin $\ft12$ & $(1,0,0)$ & $(\ft12,\ft12,-\ft12)$ & $\ft92$ & 16 \\ \hline
Vector & $(0,1,0)$ & $(1,0,0)$ & 5 & 10 \\ \hline
3-form & $(2,0,0)$ & $(1,1,-1)$ & 5 & 5 \\ \hline
Spin $\ft32$ & $(1,1,0)$ & $(\ft32,\ft12,-\ft12)$ & $\ft{11}{2}$ & 4 \\ \hline
Spin 2 & $(0,2,0)$ & $(2,0,0)$ & 6 & 1 \\ \hline
\end{tabular}}
\bigskip

\noindent{\bf Table 3:} The fields of the $D=7$ massless supergravity
multiplet, with their $SU(4)$ little-group representations in 
Dynkin and highest-weight labelling, their $E_0$ values, and their
$SO(5)$ R-symmetry representations.
\bigskip\bigskip

   Reading off the relevant partition functions from (\ref{ads7part}), 
we can then take the inverse Laplace transforms to get the energies $E_k$ 
and degeneracies $d_k$ for the fields in the supergravity multiplet.  Then
we evaluate the Casimir energies for each field.  The results are given
in Table 4 below

\bigskip\bigskip
\centerline{
\begin{tabular}{|c|c|c|c|c|} \hline
Field & Partition function & Degeneracies $d_k$ & 
 $k$ range & Casimir energy \\ \hline \hline
Scalar & $\fft{s^4}{(1-s)^6}$  & $\ft1{120} (k-3)(k-2)(k-1)k(k+1)$
 & $k\ge 4$  & $\fft{31}{120960}$ \\ \hline
Spin $\ft12$ & $\fft{4s^{9/2}}{(1-s)^6}$  & $\ft1{960} 
    (2k-7)(2k-5)(2k-3)(2k-1)(2k+1)$
 & $k\ge \ft92$  & $-\fft{1021}{322560}$ \\ \hline
Vector & $\fft{(6-s)s^5}{(1-s)^6}$  & $\ft1{24} (k-4)(k-3)(k-2)(k-1)(k+1)$
 & $k\ge 5$  & $-\fft{39}{896}$ \\ \hline
3-form & $\fft{10s^5}{(1-s)^6}$  & $\ft1{12} (k-4)(k-3)(k-2)(k-1)k$
 & $k\ge 5$  & $-\fft{95}{6048}$ \\ \hline
Spin $\ft32$ & $\fft{4(5-s)s^{11/2}}{(1-s)^6}$  & $\ft1{480} 
    (2k-9)(2k-7)(2k-5)(2k-3)(4k+3)$
 & $k\ge \ft{11}2$  & $\fft{15293}{17920}$ \\ \hline
Spin 2 & $\fft{2(10-3s)s^6}{(1-s)^6}$  & $\ft1{60}(k-5)(k-4)(k-3)(k-2)(7k+8)$
 & $k\ge 6$  & $-\fft{4143}{1120}$ \\ \hline
\end{tabular}}
\bigskip

\noindent{\bf Table 4:} The partition functions, degeneracies and Casimir
energies for each field of the types listed.  In all cases the energies
are given by $E_k=k$, with the starting values of $k$ as given in the table.

\bigskip\bigskip

    It can be seen from Table 3 and Table 4 that in this seven-dimensional 
case, the lowest energies $E_0$ for each type of field are always the
minimum allowed in each case, and so there is no need to shift 
upwards by integers.  Taking the sum of the individual Casimir energies,
weighted by the numbers for each field, we find that the total Casimir
energy for the massless supergravity multiplet is given by
\be 
E_c = -\fft{325}{384}\,.
\ee
This can be compared with the total Casimir energy for the $(2,0)$ 
antisymmetric tensor multiplet in the boundary CFT$_6$ field theory,
which has \cite{gibpopper3} 
\be
E_c = -\fft{25}{384}\,.
\ee
The Casimir energy for the bulk theory is larger in magnitude 
by a factor of 13.

\subsection{Rotating quantum fields in AdS$_4$}

   A new feature, which enters when considering rigidly-rotating 
quantum fields, is the appearance of a velocity of light surface (VLS), 
on which the rotation speed equals the speed of light.  If the system
extends beyond the VLS, one expects that the partition functions will
exhibit singularities.  

To see this,  consider AdS$_4$, for 
which the metric is
\be
ds^2 = -(1+ r^2) dt^2 + \fft{dr^2}{1+r^2} + r^2 (d\theta^2 + \sin^2\theta
   d\phi^2)\,,
\ee
where we take $\Lambda=-3$ for simplicity.  If ${\bf
K}$ is given by (\ref{tlkv}) then  ${\bf K}$
is everywhere timelike, as long as $\Omega^2 <1$, and so there is no VLS.  
However if $\Omega^2 >1 $
then there is a VLS.  Thus one expects a singularity in the partition
function as $\Omega^2
\rightarrow 1$ from below.  In fact it should be harder and harder to
rotate the system as $\Omega$ tends to this limiting value.  

   To see this reflected in the partition function, we begin by defining
\be
x= e^{-\ft12 \beta}\,,\qquad y= e^{-\ft12 \beta \Omega}\,.
\ee
The expected singularities should therefore
arise as $xy\rightarrow 1$ or $x/y \rightarrow 1$. To check this idea
one must calculate the one-particle partition functions for the
representations of $SO(3,2)$. They have been given by Flato and
Fronsdal \cite{FlatoFronsdal}:
\ben Y(\beta,
\Omega)= \sum_E \sum _{j} \sum_{j_3} x^{2E} y^{2j_3}= 
\sum_E \sum_j n(E,j) \chi_j(y),
\een 
where the $SU(2)$ rotor partition function $\chi_j(y)$ is given
by 
\ben \chi_j(y)= {(y^{2j+1} -y^{-2j-1} ) \over y-y^{-1} }.  
\een
For
scalars in the $D(1,0)$ representation 
\ben Y_{D(1,0)} = x^{-2}
Y_{D(2,0) }, 
\een 
whilst for scalars in the $D(2,0)$ representation
\ben 
Y= \fft{ x }{( x - x^{-1} ) ( xy -  (xy)^{-1}) ( y x^{-1}
-  x y^{-1})}\,.\label{d20om}  
\een

    The massless representations $ {\rm D(s+1,s)}$ have
\ben
Y={1 \over (x-x^{-1})(y-y^{-1})} \Big[
 {(xy)^{2s} \over xy -(xy)^{-1} } -{ (x/y)^{2s} \over (x/y) -(y/ x) }
\Big]\,.
\een
The singletons are given by the ``Di" representation $ D(1, { 1\over 2})$
\ben
Y(\beta, \Omega)= { (y +y ^{-1}) \over  (xy-1/xy) (x/y -y/x)} ,
\een
and the ``Rac" representation
${\rm D({ 1\over 2}, 0)}$
\ben
Y(\beta, \Omega)= { (x+x^{-1}) \over (xy-1/xy) (x/y -y/x) }\,.
\een
The expected singularities are clearly visible in these expressions. 

   It is straightforward to see from numerical studies that the 
Bekenstein bound is more and more strongly violated, at low temperatures,
as the angular velocity $\Omega$ approaches unity.  Consider, for example, a
massless scalar field in the $D(2,0)$ representation, for which the 
single-particle partition function is given by (\ref{d20om}).  If we
let $\Omega=1-\ep$, then we find
\be
Y(\beta,1-\ep) = \fft{e^\beta\, \ep^{-1}}{\beta\, (e^\beta-1)^2\, (e^\beta+1)}
   + \fft{e^\beta(e^{2\beta} +1)}{2(e^\beta-1)^3\, (e^\beta+1)^2} +
       {\cal O}(\ep)\,.
\ee
From this we find that $L(\beta) \equiv E(\beta)-S(\beta)/\pi$ as
always attains its minimum value at $\beta=\pi$, and this is given by
\be
L_{\rm min}= - \fft{0.000198216}{\ep} -0.000312661 + {\cal O}(\ep)\,.
\ee
Thus by taking $\ep$ sufficiently small, we can achieve an arbitrarily 
large violation of the Bekenstein bound.  The situation for massless 
fields of other spins is similar.

\section{Matrix Valued Fields}

   So far, we have been calculating $Z(\beta) $ using the standard
rules for free bosons and fermions. The result is a violation of
Bekenstein's bound and certainly no Hagedorn temperature.

     However, we must take into account the fact that on $S^3 \times
 {\Bbb R}$ the fields can be matrix-valued \footnote{For ${\cal N} =4
 $ Yang-Mills theory, for example, they all transform according to to
 the adjoint of $SU(N) $.}  and thus one must consider single and
 multi-trace partition functions. For very large, strictly infinite
 matrices, and for a free theory this has been solved in
 \cite{Sundborg,Polyakov,rams}

     The answers for bosons (in the ``blind'' case)  are
\ben
Y_{\rm \,Single 
\,\,Trace} (\beta)  =- \sum^\infty  _{n=1} {\phi(n) \over n} \log 
(1-Y(n\beta))  
\een
where $\phi(n)$ is the number of integers no larger than $n$ which are
relatively prime to it 
and
\ben 
\log Z_{\rm \, Multi\,\, Trace} (\beta) = - \sum^\infty_{n=1} 
\log (1-Y(n\beta)) \label{zmt}
\een
or 
\ben
Z_{\rm \, Multi\,\, Trace} (\beta)= \prod^\infty _{n=1}\, { 1 \over
1-Y(n\beta )} \,. 
\een
This is to be contrasted with
\ben
Z_B(\beta) = \prod^\infty _{n=1}    \,   e^{ { 1 \over n} Y(n\beta) } \,. 
\een
There are similar results for fermions and extensions to include chemical
potentials. Note that using M\"obius inversion
one may pass freely between $ Z_{\rm \, Multi\,\, Trace} (\beta)$
and $Y(\beta)$.

    For a quantum field theory,there can be no Hagedorn-like behaviour, 
but using infinite dimensional  matrix valued fields this can happen
Indeed, one now finds that there is a Hagedorn temperature $T_H=1/\beta_H$,
at which the partition function $Z(\beta)$   blows up 
which is located at  $\beta =\beta_H $ where 
$Y(\beta_H)=1$
In the context of the AdS/CFT correspondence,  this Hagedorn transition
has been associated with the Hawking-Page transition. 

   It is interesting to note that the Bekenstein bound can still be
violated if we make use of the multi-trace partition function rather
than the standard ones we discussed in previous sections.  For example,
let us consider the case of the multi-trace partition function for
scalars in $\R\times S^3$, for which $Y(b)$ is given in (\ref{s3energies}).
From (\ref{zmt}) and the standard expressions $E= -\del/\del\beta \log Z$,
$S= (1-\beta \del/\del\beta)\log Z$ for the energy and entropy,
we find that the function $f(\beta)= E(\beta) -S(\beta)/\pi$ has
a minimum at $\beta=\pi$, and $f(\pi)$ is negative; we have the 
Bekenstein-violating result that
\be
E(\pi) -\fft{S(\pi)}{\pi} \approx -0.0174463\,.\label{xyz}
\ee
Inclusion of the Casimir energy contribution $E_c=1/240$ is insufficient
to turn (\ref{xyz}) positive.
Note that the Hagedorn temperature is given by $\beta_H \approx 
1.25606$ in this example.

\section{Partition Functions and Temperature Inversion}

   There has been considerable interest in the behaviour under 
temperature inversion of thermal
correlation functions in conformally-invariant quantum field theories.
(See, for example, \cite{niktod} for a recent discussion, 
with references to earlier work.)  
Thermal correlators are always periodic or antiperiodic
in imaginary time, and for conformally-invariant fields on $\R\times S^{n-2}$, 
or on AdS$_{n}$, they are typically also periodic or antiperiodic in
real time.  Thus one expects elliptic functions and modular behaviour
to arise.  This has been seen to happen, for free fields at least, on
$\R\times S^{n-2}$ \cite{DowkerKirsten,niktod} and
on AdS$_4$ \cite{AllenFolacciGibbons}.  In what follows, we shall 
discuss the behaviour of the energies under temperature inversion in
certain $\R\times S^{n-2}$ and AdS$_n$ examples.   The results for
$\R\times S^{n-2}$ have been discussed previously in the literature, but
we believe that our results for AdS$_4$ are new.

\subsection{Partition functions for $\R\times S^3$}

    Here we calculate the free-field partition functions for the 
scalar, vector and spinor fields that constitute the ${\cal N}=1$
supersymmetric Yang-Mills theory on the boundary of AdS$_5$.  Our
focus will be on the relations between the high-temperature and
low-temperature limits of the free energies, which are obtained
from the multi-particle partition functions via
\be
E(\beta) = -\fft{\del\log Z}{\del\beta}\,.
\ee
First, we consider the
case where the angular momentum vanishes.

   The energies $E_n$ and degeneracies $d_n$ for the relevant
massless fields are given by
\bea
\hbox{Scalar}: && E_n= n+1\,,\qquad d_n= (n+1)^2\,,\nn\\
\hbox{Vector}: && E_n= n+2\,,\qquad d_n= 2(n+2)^2 -2\,,\nn\\
\hbox{Spinor}: && E_n = n+\ft32\,,\qquad d_n= 2(n+\ft32)^2 -\ft12\,,
\label{s3energies}
\eea
where in each case, $n$ ranges over the non-negative integers.
From these, we obtain the single-particle partition functions $Y(\beta)=
 \sum_{n\ge 0} d_n\, e^{-\beta E_n}$, given by
\bea
\hbox{Scalar}: && Y(\beta) = \fft{e^\beta\, (e^\beta +1)}{(e^\beta-1)^3}\,,
\nn\\
\hbox{Vector}: && Y(\beta) = \fft{2 (3e^\beta -1)}{(e^\beta-1)^3}\,,
\nn\\
\hbox{Spinor}: && Y(\beta) = \fft{4 e^{\ft32\beta} }{(e^\beta-1)^3}\,.
\eea
 
   The multi-particle partition functions can be calculated from these using
(\ref{bosonmp}) and (\ref{fermionmp}).  Alternatively, and completely
equivalently, they can be expressed directly via
\bea
\hbox{Boson}: &&\log Z_B = -\sum_{n\ge0} d_n \, \log(1-e^{-\beta E_n})\,,\nn\\
\hbox{Fermion}: &&\log Z_F = \sum_{n\ge0} d_n \, \log(1+ e^{-\beta E_n})\,,
\eea
whence we obtain the free energies
\bea
\hbox{Boson}: && E(\beta) = \sum_{n\ge0}\fft{d_n E_n}{e^{\beta E_n} -1}
     \,,\nn\\
\hbox{Fermion}: &&E(\beta) = \sum_{n\ge0} \fft{d_n E_n}{e^{\beta E_n}
                              +1}\,.\label{s3free}
\eea

\subsubsection{Temperature-inversion relations}
   
     The behaviour under temperature inversion, i.e.
under  $\beta \rightarrow \beta^{-1}$, may be determined
from the behaviour of the integrands in (\ref{bint}) and (\ref{fint})
under $s\rightarrow -s$.
This is particularly easy to investigate in cases where
$\zeta_H(s)$ is known explicitly.  An alternative procedure, as discussed 
in \cite{DowkerKirsten}, is to make use of the Ramanujan formulae \cite{raman}
\bea
\mu^p \sum_{n\ge 1} \fft{n^{2p -1} }{e^{2\mu n} -1} 
   -(-\td\mu)^t\, \sum_{n\ge 1} \fft{ n^{2p-1} }{e^{2\td\mu n} -1} &=&
   [\mu^p - (-\td\mu)^p]\, \fft{B_{2p} }{4p}\,,\label{ramab}\\
\mu^p \sum_{n\ge 0} \fft{(2n+1)^{2p -1}}{e^{(2n+1)\mu} +1}
   -(-\td\mu)^p\, \sum_{n\ge1} \fft{ (2n+1)^{2p-1}}{e^{(2n+1)\td\mu}+1} &=&
   [\mu^p - (-\td\mu)^p]\, (2^{2p-1}-1)\, \fft{B_{2p} }{4p}\,,\label{ramaf}
\eea
for the bosonic and fermionic sums, respectively, where
\be
\mu\, \td\mu = \pi^2\,,
\ee
$p$ is a positive integer, and $B_n$ denotes the $n$'th Bernoulli number.
We take $\mu=\ft12 \beta$.  

\subsubsection{Scalar field}

   For the scalar field, one obtains
\be
E(\beta) = \Big(\fft{2\pi}{\beta}\Big)^4\, E\Big(\fft{4\pi^2}{\beta}\Big)
    + \ft18 \Big[1- \Big(\fft{2\pi}{\beta}\Big)^4\Big]\, B_4 \,.
\ee
 From (\ref{Cas}), we see that the Casimir energy is
\be
E_c = \ft12 \zeta(-3) = -\ft18 B_4 = \ft1{240}\,,
\ee
where $\zeta(s)=\sum_{n\ge 1} n^{-s}$ is the Riemann zeta function.
It follows that the total energy $E_{\rm tot}(\beta)$ of the scalar
field, defined by
\be
E_{\rm tot}(\beta) \equiv E(\beta) + E_c\,,
\ee
satisfies the inversion relation 
\be
E_{\rm tot}(\beta) = \Big(\fft{2\pi}{\beta}\Big)^4\, 
       E_{\rm tot}\Big(\fft{4\pi^2}{\beta}\Big)\,.
\label{scalarinvert}
\ee

   Recalling that we have set the AdS radius to $l=1$ in our calculations,
it can be seen that (\ref{scalarinvert}) implies a temperature inversion
symmetry under $T\longrightarrow l^2/(4\pi^2 T)$. 
 
\subsubsection{Vector field}

   The total energy of the scalar field in $\R\times S^3$ obeys the 
elegant inversion formula (\ref{scalarinvert}).  It turns out that
for fields of non-zero spin, the analogous inversion formulae are not
so elegant, but they still take rather simple forms.  It is is helpful
first to define functions
\be  
f_p(\beta) \equiv \sum_{n\ge 1} \fft{n^{2p-1}}{e^{n\beta} -1}\,.
\label{ftdef}
\ee
Using (\ref{ramab}), these functions satisfy the inversion relations
\be
f_p(\beta) = (-1)^p\, \Big(\fft{2\pi}{\beta}\Big)^{2t}\, 
    f_t\Big(\fft{4\pi^2}{\beta}\Big) + 
 \Big[1- (-1)^p \Big(\fft{2\pi}{\beta}\Big)^{2p}\Big]\,\fft{B_{2p}}{4p}\,.
\label{ftinvert}
\ee

   From (\ref{s3energies}) and (\ref{s3free}), it then follows that the 
free energy for the vector field is given by
\be
E(\beta) = 2[ f_2(\beta) - f_1(\beta)]\,.
\ee
The Casimir energy calculated using (\ref{Cas}) is
\be
E_c = \zeta(-3) -\zeta(-1) = -\ft14 B_4 +\ft12 B_2 = 
       \ft1{120} + \ft1{12}= \ft{11}{120}\,,
\ee
and hence we see from (\ref{ftinvert}) that the total energy 
$E_{\rm tot}(\beta)=E_c + E(\beta)$ for the vector field on $\R\times S^3$
satisfies the inversion relation
\be
E_{\rm tot}(\beta) = 2\Big(\fft{2\pi}{\beta}\Big)^4\, \Big[
       f_2\Big(\fft{4\pi^2}{\beta}\Big) +\ft1{240}\Big] +
   2\Big(\fft{2\pi}{\beta}\Big)^2\, \Big[
       f_1\Big(\fft{4\pi^2}{\beta}\Big) +\ft1{24}\Big]\,.\label{vectorinvert}
\ee

\subsubsection{Spinor field}

    Again, one finds that the total energy for a spin-$\ft12$ Weyl fermion
in $\R\times S^3$ does not satisfy as simple an inversion relation as the
scalar field in (\ref{scalarinvert}), but a more complicated relation that is
similar to the vector-field relation (\ref{vectorinvert}).  It is helpful 
first to define
\be
g_p(\beta) \equiv \sum_{n\ge 0} \fft{(2n+1)^{2p-1}}{e^{(n+\ft12)\beta} +1}\,,
\label{gtdef}
\ee
Using (\ref{ramaf}), these functions satisfy the inversion relations
\be
g_p(\beta) = (-1)^p\, \Big(\fft{2\pi}{\beta}\Big)^{2p}\,
    g_p\Big(\fft{4\pi^2}{\beta}\Big) +
 \Big[1- (-1)^p \Big(\fft{2\pi}{\beta}\Big)^{2t}\Big](2^{2p-1} -1)\,
 \fft{B_{2t}}{4p}\,.
\label{gtinvert}
\ee

    It follows from (\ref{s3energies}) and (\ref{s3free}) that the 
free energy for a spinor field is given by
\be
E(\beta) = \ft14 [ g_2(\beta) - g_1(\beta)]\,.
\ee
 From (\ref{s3energies}) and (\ref{Cas}) it follows that the Casimir
energy for the spinor field is given by
\be
E_c = -\zeta(-3,\ft12) + \ft14 \zeta(-1,\ft12) = \ft{7}{960} + \ft1{96}=
     \ft{17}{960}\,,
\ee
and hence using (\ref{gtinvert}) we find that the total energy
$E_{\rm tot}(\beta)=E_c + E(\beta)$ obeys the inversion relation
\be
E_{\rm tot}(\beta) = 
\ft14 \Big(\fft{2\pi}{\beta}\Big)^4\, \Big[
       g_2\Big(\fft{4\pi^2}{\beta}\Big) -\ft7{240}\Big] +
   \ft14\Big(\fft{2\pi}{\beta}\Big)^2\, \Big[
       g_1\Big(\fft{4\pi^2}{\beta}\Big) -\ft1{24}\Big]\,.\label{spinorinvert}
\ee

\subsection{Partition functions for $\R\times S^5$}

   In this six-dimensional boundary case, the relevant fields fill out
a $(2,0)$ supermultiplet, comprising one vector, four spinors, and five 
self-dual tensor multiplets.  Since the analysis is very similar to that
in the $\R\times S^3$ boundary theory that we discussed in the previous
section, where we shall be rather brief, and just focus on the results.

   The energies and degeneracies for the three fields are given by
\bea
\hbox{Scalar}: && E_n = n+2\,,\qquad d_n = \ft1{12} (n+2)^4 - 
                            \ft1{12} (n+2)^2\,,\nn\\
\hbox{Tensor}: && E_n = n+3\,,\qquad d_n = \ft1{4} (n+3)^4 -\ft54
                     (n+3)^2 + 1\,,\nn\\
\hbox{Spinor}: && E_n = n+\ft52\,,\qquad d_n = \ft1{3} (n+\ft52)^4 -
                            \ft5{6} (n+ \ft52)^2 + 
                                     \ft3{16}\,,\label{s5energies}
\eea
where $n\ge0$, implying that the single-particle partition functions are
\bea
\hbox{Scalar}: && Y(\beta) = \fft{e^{2\beta}(e^\beta +1)}{(e^\beta-1)^5}
\,,\nn\\
\hbox{Tensor}: && Y(\beta) = \fft{10e^{2\beta} -5e^\beta +1}{(e^\beta-1)^5}
\,,\nn\\
\hbox{Spinor}: && Y(\beta) = \fft{8e^{\ft52\beta} }{(e^\beta-1)^5}\,.
\eea

    In terms of the functions $f_p(\beta)$ and $g_p(\beta)$ defined in 
(\ref{ftdef}) and (\ref{gtdef}), the free energies turn out to be
given by
\bea
\hbox{Scalar}: && E(\beta) = \ft1{12} [f_3(\beta) - f_2(\beta)]\,,\nn\\
\hbox{Tensor}: && E(\beta) = \ft1{4} [f_3(\beta) - 5f_2(\beta)
                                  + 4 f_1(\beta) ]\,,\nn\\
\hbox{Spinor}: && E(\beta) = \ft1{96} [g_3(\beta) - 10g_2(\beta)
                      + 9 g_1(\beta)]\,.\label{s5free}
\eea
The Casimir energies, calculated from (\ref{Cas}) and (\ref{s5energies}),
are given by
\bea
\hbox{Scalar}:&& E_c= \ft1{24} \zeta(-5) - \ft1{24}\zeta(-3) =
                        -\ft{31}{60480}\,,\nn\\
\hbox{Tensor}:&& E_c= \ft1{8} \zeta(-5) - \ft5{8}\zeta(-3) 
              + \ft12 \zeta(-1) =
                        -\ft{191}{4032}\,,\nn\\
\hbox{Spinor}:&& E_c= -\ft1{6} \zeta(-5,\ft12) + \ft5{12}\zeta(-3,\ft12)
              - \ft3{32} \zeta_(-1,\ft12) =
                        -\ft{367}{48384}\,.
\eea

    As we saw in the previous four-dimensional case, here again the 
Casimir energies coincide with the (negatives of the) 
$\beta$-independent terms coming from the right-hand sides of the 
temperature-inversion relations (\ref{ftinvert}) and (\ref{gtinvert}),
once these are assembled into the combinations occurring in (\ref{s5free}).
This implies that the total energies $E_{\rm tot}(\beta)= E_c + E(\beta)$
satisfy the following temperature-inversion relations:
\bea
\hbox{Scalar}:&& E_{\rm tot}(\beta) = -\ft1{12}\Big(\fft{2\pi}{\beta}\Big)^6
  \Big[ f_3\Big(\fft{4\pi^2}{\beta}\Big)  -\ft1{504}\Big] 
   -\ft1{12}\Big(\fft{2\pi}{\beta}\Big)^4
  \Big[ f_2\Big(\fft{4\pi^2}{\beta}\Big)  +\ft1{240}\Big]\,,\nn\\
\hbox{Tensor}:&& E_{\rm tot}(\beta) = -\ft1{4}\Big(\fft{2\pi}{\beta}\Big)^6
  \Big[ f_3\Big(\fft{4\pi^2}{\beta}\Big)  -\ft1{504}\Big]
   -\ft5{4}\Big(\fft{2\pi}{\beta}\Big)^4
  \Big[ f_2\Big(\fft{4\pi^2}{\beta}\Big)  +\ft1{240}\Big]
\nn\\
&&\qquad\qquad\qquad  -\Big(\fft{2\pi}{\beta}\Big)^2
  \Big[ f_1\Big(\fft{4\pi^2}{\beta}\Big)  -\ft1{24}\Big]
\,,\nn\\
\hbox{Spinor}:&& E_{\rm tot}(\beta) = -\ft1{96}\Big(\fft{2\pi}{\beta}\Big)^6
  \Big[ g_3\Big(\fft{4\pi^2}{\beta}\Big)  -\ft{31}{504}\Big]
   -\ft5{48}\Big(\fft{2\pi}{\beta}\Big)^4
  \Big[ g_2\Big(\fft{4\pi^2}{\beta}\Big)  +\ft7{24}\Big]
  \nn\\
 &&\qquad\qquad\qquad +\ft{3}{32} \Big(\fft{2\pi}{\beta}\Big)^2
  \Big[ g_1\Big(\fft{4\pi^2}{\beta}\Big)  +\ft3{8}\Big]
\,.
\eea

\subsection{Temperature-inversion formulae in AdS$_4$}

   The energies $E_n$ and degeneracies $d_n$ for the modes in AdS$_4$ 
can conveniently be read off from (\ref{rhoads4}) and (\ref{rhoads42}).
Thus we have
\bea
D(1,0):&& E_n = n\,,\qquad d_n = \ft12 n(n+1)\,,\qquad n\ge 1\,,\nn\\
D(2,0):&& E_n = n\,,\qquad d_n = \ft12 n(n-1)\,,\qquad n\ge 2\,,\nn\\
D(s+1,s):&& E_n = n\,,\qquad d_n = n^2-s^2    \,,\qquad n\ge s\,,\nn\\
D(1,\ft12):&& E_n=n\,,\qquad d_n= n\,,\qquad n\ge 1\,,\nn\\
D(\ft12,0):&& E_n=n\,,\qquad d_n= n\,,\qquad n\ge \ft12 \,,
\eea
where in each case $n$ increases in integer step.

   The free energies can be calculated using the same formulae 
(\ref{s3free}) that we used when studying the case of $\R\times S^3$.  
Thus we shall have
\bea
E_{(1,0)}(\beta) &=& \ft12 \sum_{n\ge 1} 
           \fft{n^2(n+1)}{e^{n\beta} -1}\,,\qquad
E_{(2,0)}(\beta) = \ft12 \sum_{n\ge 1}\fft{n^2(n-1)}{e^{n\beta} -1}\,,\nn\\
E_{(s+1,s)}(\beta) &=& 
  \sum_{n\ge s} \fft{n(n^2-s^2)}{e^{n\beta} -1}\,,\qquad s\in \Z \nn\\
E_{(s+1,s)}(\beta) &=& 
  \sum_{n\ge s} \fft{n(n^2-s^2)}{e^{n\beta} +1}\,,\qquad s\in \Z+\ft12 \nn\\
E_{(1,\ft12)}(\beta)&=& \sum_{n\ge 1} \fft{ n^2}{e^{n\beta} +1}\,,\qquad
E_{(\ft12,0)}(\beta)= \sum_{n\ge \ft12} \fft{ n^2}{e^{n\beta} -1}\,.
\eea

   Let us consider scalars first.  It is evident from the Ramanujan formula
(\ref{ramab}) that, since $p$ must be an integer there, we cannot obtain
a temperature-inversion formula for the $D(1,0)$ or $D(2,0)$ scalars
separately.  However, if we add up the two, we get
\be
E_{\rm scal}(\beta) = \sum_{n\ge 1} 
           \fft{n^3}{e^{n\beta} -1}\,,
\ee
which does fall into the category covered by (\ref{ramab}).  In fact, from
(\ref{ftdef}) and (\ref{ftinvert}), and noting from (\ref{ads4casimirs})
that the total scalar Casimir energy is $\ft1{240}$, we shall have for 
the total scalar energy $E_{\rm tot}(\beta)\equiv E_{\rm scal}(\beta) +
E_c$ the same total energy, and temperature-inversion formula
\be
E_{\rm tot}(\beta) = \Big(\fft{2\pi}{\beta}\Big)^4\, 
        E_{\rm tot}(\fft{4\pi^2}{\beta})\,,
\ee
as we obtained for scalars in $\R\times S^3$.

\section{Conclusions}

     In this paper, motivated by our earlier work on black holes and the
cosmic censorship bound in AdS \cite{GibbonsPerryPope}, we have investigated
a precise formulation of the Bekenstein bound four quantum fields in 
AdS$_n$.  We have given it a new formulation in terms of the function $L\equiv
 E - (n-2) S/(2\pi l)$, where $l$ is the AdS radius.  If the specific
heat is positive, the function $L$ has a unique minimum at the temperature
$T=T_L\equiv (n-2)/(2\pi l)$, at which $L=F(T_L)$, where $F$ is
the Helmholtz free energy.  Thus the Bekenstein bound is satisfied
if and only if $F(T_L)\ge0$.  Interestingly, the marginal case 
corresponds to the free energy having the same value as that of the
vacuum, as it does in the case of the Hawking-Page phase transition.
Although we found previously that the Bekenstein bound was satisfied for
all known black holes, we are able to exhibit violations of the bound
for free quantum fields of various spins in AdS, including in particular
those which come from supermultiplets.  We do this by calculating the
bulk entropies and energies in AdS$_4$, AdS$_5$ and AdS$_7$.  
We have also examined rotating quantum fields in AdS$_4$, where we
find the expected divergence in the partition function as the rotation
rate tends to its maximum value.  Violations of the Bekenstein bound  
can be arbitrarily large as this limit is approached.

   A summary of the status of the conjectured Bekenstein bound is as
follows.  As has been observed previously, it is trivially invalid if one
uses the Boltzmann definition of entropy in the microcanonical ensemble,
because of the species problem.  Furthermore, inclusion of the Casimir
energy cannot always rescue it, since sometimes the Casimir energy is
negative.  In this paper, we have resolved a previously difficulty with
the bound, which is the precise definition of the radius $R$, and the
problem of dealing with the boundary conditions on the surface of the box.
These problems, which plagued previous discussions, were avoided in a
simple and natural way by working in anti-de Sitter spacetime.  We 
considered the Gibbs definition of entropy in the canonical and 
grand canonical ensembles, and showed that we could obtain violations
of the Bekenstein bound in this well-defined situation.  Thus it appears
that there is no definition of entropy for which a rigorous Bekenstein
bound holds.

    It has recently been realised that the statistics of matrix-valued
fields in Yang-Mills theory are non-conventional, and we have calculated
some entropies and energies for bulk fields in AdS using these 
non-conventional statistics.  Again, we find that the Bekenstein 
bound may be violated.

   A topic of some interest in AdS$_n$ and in $\R\times S^{n-2}$ is the
issue of a possible symmetry of thermodynamic quantities
under temperature inversion \cite{DowkerKirsten,niktod}.  We have 
investigated this in $\R\times S^3$, $\R\times S^5$ and AdS$_4$.  The
results for $\R\times S^{n-2}$ have appeared previously, but our results
for AdS$_4$ are new.  

   String theory has a T-duality symmetry which implies that amplitudes
transform covariantly under $R\longrightarrow \alpha'/R$, where $\alpha'$
is the inverse string tension and $R$ is the radius of a Kaluza-Klein
circle.  Various authors \cite{chau,dien1,dien2} have consequently speculated
that there should be a temperature inversion symmetry in string theory, 
with $T\longrightarrow \alpha'/T$.  It should be noted, however, that
the temperature inversion we consider, where 
$T\longrightarrow l^2/(4\pi^2 T)$, is distinct from the type of temperature
inversion envisaged in string theory.

    Finally, in the appendix, we give what we believe to be a novel way
of obtaining the one-particle partition function from the many-particle
partition function, using a version of the M\"obius inversion 
formula.

\section*{Acknowledgement}

   We should like to thank Francis Dolan for helpful conversations and
for sharing with us some of his results prior to publication.  
C.N.P. is grateful
to the DAMTP Relativity and Gravitation group, and the Centre for
Theoretical Cosmology at the CMS in Cambridge, for
hospitality during the course of this work.

\appendix

\section{Multi-Particle to Single-Particle Partition Functions}

   As is well known, and we discussed in section \ref{pfsec}, one constructs
the multi-particle partition function $\widetilde Y(\beta)=\log Z$ for a 
system of non-interacting particles whose single-particle partition 
function is $Y(\beta)$ according to the formulae
\bea
\hbox{Bosons}: && \widetilde Y(\beta) = \sum_{n\ge 1} 
               \fft1{n}\, Y(n\beta)\,,\label{bosonmp}\\
\hbox{Fermions}: && \widetilde Y(\beta) = -\sum_{n\ge 1}
               \fft{(-1)^{n}}{n}\, Y(n\beta)\,.\label{fermionmp}
\eea
What appears to be less well known is that one can invert this construction,
and express the single-particle partition functions in terms of the
multi-particle partition functions.  

   For the bosonic case, the inversion can be performed by using 
the M\"obius function $\mu(n)$, which is defined for positive integers
$n$ as follows.  
If $n$ is not a square-free integer then $\mu(n)=0$, whilst $\mu(n)=(-1)^m$
if $n$ is a product of $m$ distinct primes, with $\mu(1)=1$.  It is a
standard result that if
\be
g(x) \equiv \sum_{n\ge 1} f(n x)\,,\label{gfromf}
\ee
then
\be
f(x) = \sum_{n\ge 1} \mu(n)\, g(nx)\,,\label{ffromg}
\ee
where $f(x)$ is an arbitrary function restricted only by the 
requirement that the sums converge.  From this, if follows that
for bosons we may invert (\ref{bosonmp}) to express the single-particle
partition function $Y(\beta)$ in terms of the multi-particle 
partition function $\widetilde Y(\beta)$ as
\be
Y(\beta) = \sum_{n\ge 1} \fft{\mu(n)}{n}\, \widetilde Y(n\beta)\,.
\ee

   In the fermionic case, we may again seek an inversion of the form
\be
Y(\beta) = \sum_{n\ge 1}
                f_n\,\widetilde Y(n\beta)\,.\label{fermionsp}
\ee
We find that the coefficients $f_n$ in this expansion are given
as follows.  We first express $n$ as 
\bea
n= 2^s\, \prod_{i=1}^m p_i^{c_i}\,,
\eea
where $p_i$ denotes the prime factors in $n$ that are $\ge 3$.  
The $f_n$ are then given by $f_1=1$ and
\be
f_n = \left\{ \begin{array}{ll}
    \fft{(-1)^m}{n}\, 2^{s-1} & \hbox{if $s\ge 1$ and $c_i=1$ for all $i$}\,,\\
    \fft{(-1)^m}{n} & \hbox{if $s=0$ and $c_i=1$ for all $i$}\,,\\
     \fft12 & \hbox{if $s\ge 1$ and $c_i=0$ for all $i$}\,,\\
      0 & \hbox{otherwise}\,.\label{fdef}
\end{array} \right.
\ee
To prove this, we define
\be
G(x) = \sum_{n\ge 1} \fft1{n}\, F(nx)\,,\qquad
H(x) = \sum_{n\ge 1} \fft{(-1)^{n+1}}{n}\, F(nx)\,.
\ee
By splitting the summation in the latter into the  
terms where $n$ is even and $n$ is odd, it follows that
\be
H(x) = G(x) - G(2x)\,,
\ee
which can be iterated to give
\be
G(x) = \sum_{p\ge 0} H(2^p\, x)
\ee
since $G(x)$ is assumed to go to zero as $x$ goes to infinity.  Using the
standard result that (\ref{gfromf}) implies (\ref{ffromg}), we
therefore have
\bea
F(x) &=&  \sum_{n\ge 1, p\ge 0} \fft{\mu(n)}{n}\, H(2^p n\, x)\nn\\
&=& \sum_{n\ge 1, p\ge0} \fft{\mu(2n)}{2n}\, H(2^{p+1} n\, x)
   \ + \sum_{n\, {\rm odd}\, \ge1, p\ge0} 
       \fft{\mu(n)}{n}\, H(2^{p} n\, x)\,,
\eea
where the two terms in the second line were 
obtained by splitting the original sum over $n$ 
into the cases where $n$ is even and odd respectively.  Since $\mu(2n)=
-\mu(n)$ if $n$ is odd, whilst $\mu(2n)=0$ if $n$ is even, it follows 
that
\bea
F(x) &=&  \sum_{n\, {\rm odd}\, \ge1, p\ge0} 
       \fft{\mu(n)}{n} [H(2^p n\, x) - \ft12 H(2^{p+1} n\, x)]\nn\\
&=& \sum_{n\, {\rm odd}\, \ge 1} \fft{\mu(n)}{n}\, H( nx) \ +
  \sum_{n\, {\rm odd}\, \ge1, p\ge1} \fft{\mu(n)}{(2^p n)}\,
   2^{p-1}\, H(2^p n x)\,.
\eea
Comparing with (\ref{fermionmp}) and (\ref{fermionsp}), we see that
(\ref{fdef}) is indeed established.


\begin{thebibliography}{99}

\bm{bek} J.D. Bekenstein,
{\it A universal upper bound on the entropy to energy ratio for bounded
systems}, 
Phys. Rev. {\bf D23}, 287 (1981).

\bm{WaldUnruh} W.G. Unruh and R.M. Wald,
{\it Acceleration radiation and generalized second law of thermodynamics}, 
Phys. Rev. {\bf D25}, 942 (1982).

\bm{feynman} R.P. Feynman, {\it Statistical mechanics}, (W.A. Benjamin,
Reading, Mass., 1972). 

\bibitem{HawkingPage} S.W. Hawking and D.N. Page, {\it Thermodynamics 
of black holes in anti-de Sitter space}, Commun. Math. Phys.
{\bf 87}, 577 (1983).

\bibitem{AllenFolacciGibbons} B. Allen, A. Folacci and G.W. Gibbons,
{\it Anti-de Sitter space at finite temperature},  Phys. Lett. {\bf B 189}, 
304 (1987).
 
\bm{BreitenlohnerFreedman} P. Breitenl\"ohner and D.Z. Freedman,
{\it Positive energy
in anti-de Sitter backgrounds and gauged extended supergravity},
Phys. Lett. {\bf B139} (1984) 154.

\bm{Hawking} S.W. Hawking,
{\it The boundary conditions for gauged supergravity}, 
Phys. Lett. {\bf B126}, 175 (1983).

\bm{AllenDavis} B. Allen and S. Davis,
{\it Vacuum energy in gauged extended supergravity}, 
Phys. Lett. {\bf B124}, 353 (1983).

\bm{GibbonsNicolai} G.W. Gibbons and H. Nicolai,
{\it One loop effects on the round seven sphere}, 
Phys. Lett. {\bf B143}, 108 (1984).

\bibitem{Dowker2} J.S. Dowker, 
{\it Vacuum energy in a squashed Einstein universe}, in
``Quantum Theory of Gravity,'' (ed. S.M. Christensen, Adam Hilger, Bristol, 
1984).

\bibitem{KutasovLarsen} D. Kutasov and F. Larsen,
{\it Partition sums and entropy bounds
in weakly coupled CFT}, JHEP {\bf 0101}, 001 (2001), hep-th/0009244.

\bm{Klemm} D. Klemm, A.C. Petkou and G. Siopsis,
{\it Entropy bounds, monotonicity properties and scaling in CFTs}, 
Nucl. Phys. {\bf B601}, 380 (2001), hep-th/0101076.

\bibitem{Dowker} J.S. Dowker, 
{\it Zero modes, entropy bounds and partition functions}, 
Class. Quant. Grav. {\bf 20}, L105 (2003), hep-th/0203026.

\bm{Odd1} I. Brevik, K.A. Milton and S.D. Odintsov,
{\it Entropy bounds in $R \times S^3$ geometries}, 
Ann. Phys. (NY)  {\bf 302}, 120 (2002), hep-th/0202048.

\bm{Odd2} I. Brevik, K.A. Milton and S.D. Odintsov,
{\it Entropy bounds in spherical space}, hep-th/0210286.

\bm{GibbonsPerryPope} G.W. Gibbons, M.J. Perry and C.N. Pope,
{\it Bulk/boundary thermodynamic equivalence, and the Bekenstein and
cosmic-censorship bounds for rotating charged AdS black holes}, 
Phys. Rev. {\bf D72}, 084028 (2005), 
hep-th/0506233.

\bm{Sundborg} B. Sundborg,
{\it The Hagedorn transition, deconfinement and $N = 4$ SYM theory},
Nucl. Phys. {\bf B573}, 349 (2000), hep-th/9908001.

\bm{Polyakov} A.M. Polyakov,
{\it Gauge fields and space-time},
Int. J. Mod. Phys. {\bf A17S1}, 119 (2002), hep-th/0110196.

\bm{rams} O. Aharony, J. Marsano, S. Minwalla, K. Papadodimas and
M. Van Raamsdonk,
{\it The Hagedorn/deconfinement phase transition in weakly coupled
large $N$ gauge theories},
Adv.  Theor. Math. Phys.  {\bf 8}, 603 (2004), hep-th/0310285.


\bm{tolman} R.C. Tolman, {\it On the weight of heat, and thermal equilibrium
in general relativity}, Phys. Rev. {\bf 35}, 904 (1930).

\bm{dolan} F.A. Dolan,
{\it Character formulae and partition functions in higher 
dimensional conformal field theory}, hep-th/0508031.

\bm{guntak} M. Gunaydin and S. Takemae,
{\it Unitary supermultiplets of OSp(8*$|$4) and the\newline 
AdS$_7$/CFT$_6$ duality,}
Nucl. Phys. {\bf B578}, 405 (2000), Erratum-ibid. {\bf B697}, 399 (2004),
hep-th/9910110.

\bm{gibpopper3} G.W. Gibbons, M.J. Perry and C.N. Pope,
{\it AdS/CFT Casimir energy for rotating black holes},
Phys. Rev. Lett. {\bf 95}, 231601 (2005), hep-th/0507034.

\bibitem{FlatoFronsdal} M. Flato and C. Fronsdal,
{\it One massless particle equals two Dirac singletons},
Lett. Math. Phys. {\bf 2}, 421 (1978).

\bibitem{DowkerKirsten} J.S. Dowker and K. Kirsten,
{\it Elliptic functions and temperature inversion
symmetry on spheres}, Nucl. Phys. {\bf B638}, 405 (2002), hep-th/0205029.

\bm{niktod} N.M. Nikolov and I.T. Todorov,
{\it Elliptic thermal correlation functions and modular forms in a globally
conformal invariant QFT}, 
Rev. Math. Phys. {\bf 17}, 613 (2005).

\bm{raman} B.C. Berndt, {\it Ramanujan's notebooks}, Vol. 3, Chapter 21
(Springer 1991). 

\bm{chau} S. Chaudhuri,
{\it Finite temperature bosonic closed strings: Thermal duality and the KT
transition}, 
Phys. Rev. {\bf D65}, 066008 (2002), hep-th/0105110.


\bm{dien1} K.R. Dienes and M. Lennek,
{\it Adventures in thermal duality. I: Extracting closed-form solutions for
finite-temperature effective potentials in string theory},
Phys. Rev. {\bf D70}, 126005 (2004), hep-th/0312216.

\bm{dien2} K.R. Dienes and M. Lennek,
{\it Adventures in thermal duality. II: Towards a duality-covariant string
thermodynamics},
Phys. Rev. {\bf D70}, 126006 (2004), hep-th/0312217.

\end{thebibliography}
\end{document}